\documentclass[groupedaddress,nofootinbib, showpacs]{revtex4}  
\usepackage{graphics}
\renewcommand{\wr}{\omega_r}
\newcommand{\wb}{\omega_b}
\begin{document}

\voffset=0.2in
\title{Geometric structure of coexisting phases found in the\\
Biham-Middleton-Levine traffic model}

\author{Raissa M. D'Souza} \email{raissa@microsoft.com}
\affiliation{Microsoft Research, Redmond WA 98052} \date{September 2003}

\begin{abstract}
The Biham-Middleton-Levine traffic model is perhaps the simplest
system exhibiting phase transitions and self-organization.  Moreover,
it is an underpinning to extensive modern studies of traffic flow.
The general belief is that the system exhibits a sharp phase
transition from freely flowing to fully jammed, as a function of
initial density of cars.  However we discover intermediate stable
phases, where jams and freely flowing traffic coexist.  The geometric
structure of such phases is highly regular, with bands of free flowing
traffic intersecting at jammed wavefronts that propagate smoothly
through the space. Instead of a phase transition as a function of
density, we see bifurcation points, where intermediate phases begin
coexisting with the more conventionally known phases.  We show that
the regular geometric structure is in part a consequence of the finite
size and aspect ratio of the underlying lattice, and that for certain
aspect ratios the asymptotic intermediate phase is on a short periodic
limit cycle (the exact microscopic configuration recurs each $\tau$
timesteps, where $\tau$ is small compared to the system size).  Aside
from describing these intermediate states, which previously were
overlooked, we derive simple equations to describe the geometric
constraints, and predict their asymptotic velocities.
\end{abstract}

\maketitle

\section{Introduction}\label{sec:intro}
Modeling vehicular and Internet traffic, thereby gaining an
understanding of congestion patterns and jamming phenomena, is an
extremely relevant problem, with obvious practical ramifications. One
popular approach is the use of simple discrete cellular automata (CA)
models, which capture aspects of the dynamics of discrete vehicles or
packets. One of the most cited examples of such a CA model is the
Biham, Middleton, and Levine model (BML) of two-dimensional traffic
flow\cite{BML92}. At the time of this writing, Ref.~\cite{BML92} has
received over two-hundred citations in the scientific literature, and
it serves as a theoretical underpinning for the physicists' approach
to modeling traffic. Note that using techniques from physics to model
traffic has been a fruitful research area for more than a decade, and
continues to be.  For recent reviews see
Refs.~\cite{NagaRev2002,SchadRev2002,ChowSanSchad-Rev2000,NagelRev96}.

The BML model describes two species of ``cars'' moving on a
two-dimensional square lattice, with periodic boundary conditions.
The model is extremely simple, yet the behaviors it displays,
extraordinarily complex.  The system shows what appears to be a phase
transition from having all cars freely moving at all time steps, to
complete gridlock, where no car can ever move again.  In addition, in
all the phases, the system becomes fully correlated, forming a range
of interesting stable self-organized patterns.  It is perhaps the
simplest model where one can study both phase transitions and self
organization.  This model has recently become increasingly of interest
to the combinatorial mathematics community, as it continues to elude
rigorous theoretical analysis\cite{combinatorics-guys}.

We implement the BML model and experimentally study its behaviors.  We
discover stable intermediate states that have never been reported
before for the BML model, with highly structured geometric patterns of
wave fronts of jams moving through otherwise freely flowing
traffic. We show the geometry of these patterns arise due to the
finite size and periodic boundary conditions of the underlying
lattice.  We also show that the aspect ratio of the lattice imposes
geometric constraints which restrict the patterns, and derive simple
equations describing these geometric constraints which allow us to
calculate the asymptotic velocities.  For certain aspect ratios we can
prove that the intermediate configurations end up on a short periodic
limit cycle---the exact microscopic configurations recur each $\tau$
timesteps (where $\tau$ is small compared to the system size)---hence
these states are stable for all time.  For the other aspect ratios, we
show the intermediate states are at least metastable, lasting as long
as we could simulate them. By establishing the existence of these
intermediate states, we show that the conventional beliefs about this
model need to be reexamined.  Contrary to the evidence published
elsewhere, only on smaller spaces do we see evidence for a sharp
transition from freely flowing to fully jammed configurations as a
function of the initial density of cars, $\rho$.  Instead we observe
bifurcations as a function of $\rho$, where different phases can begin
to coexist.  The bifurcation points, the range of the windows for
phase coexistence, and the number of coexisting phases, depend on the
size and the aspect ratio of the underlying lattice.  Considering the
amount of ongoing work on this model, and its use in large scale,
complex simulations of traffic, we believe it is important to
understand these new observations.

This manuscript is organized as follows.  In Sec.~\ref{sec:review} we
review the BML model and relevant past work.  In
Sec.~\ref{sec:results} we describe our simulations and empirical
results. Section~\ref{sec:geom} contains a discussion of the kinetic
pathways, geometric constraints, and derivation of the velocities for
the intermediate states.  Finally, in Sec.~\ref{sec:disc}, we
summarize and discuss open questions and areas for further inquiry.

\section{The BML model}\label{sec:review}

Consider two species of particles ({\em i.e.}, ``cars''), east-bound
and north-bound (which we also interchangeably call ``red'' and
``blue'' respectively), which populate a two-dimensional square
lattice with periodic boundary conditions.  Each lattice site can be
in one of three states: empty, occupied by an east-bound car, or
occupied by a north-bound car. The cars are initially distributed
uniformly at random along the lattice sites, with spatial density,
$\rho$ (usually taken to be the same for both north and east-bound
cars).  The discrete time dynamics has two phases.  On even steps, all
east-bound cars synchronously attempt to advance one lattice site
towards the east. If the site eastward of a car is currently empty, it
advances. Otherwise, it remains stationary (even if the eastward site
is to become empty during the current time step).  On odd time steps,
the north-bound cars follow the analogous dynamics, only attempting to
advance to the north-bound site.  The dynamics is fully deterministic.
The only randomness is in the initial condition.  Furthermore, the
dynamics conserves cars, and does not allow for an east-bound car to
change its row, nor for a north-bound car to change its column. So on
a $L\times L$ lattice, there are $2L$ conservation laws.  

\begin{figure}[tb]
{\hfill\resizebox{2in}{!}{\includegraphics{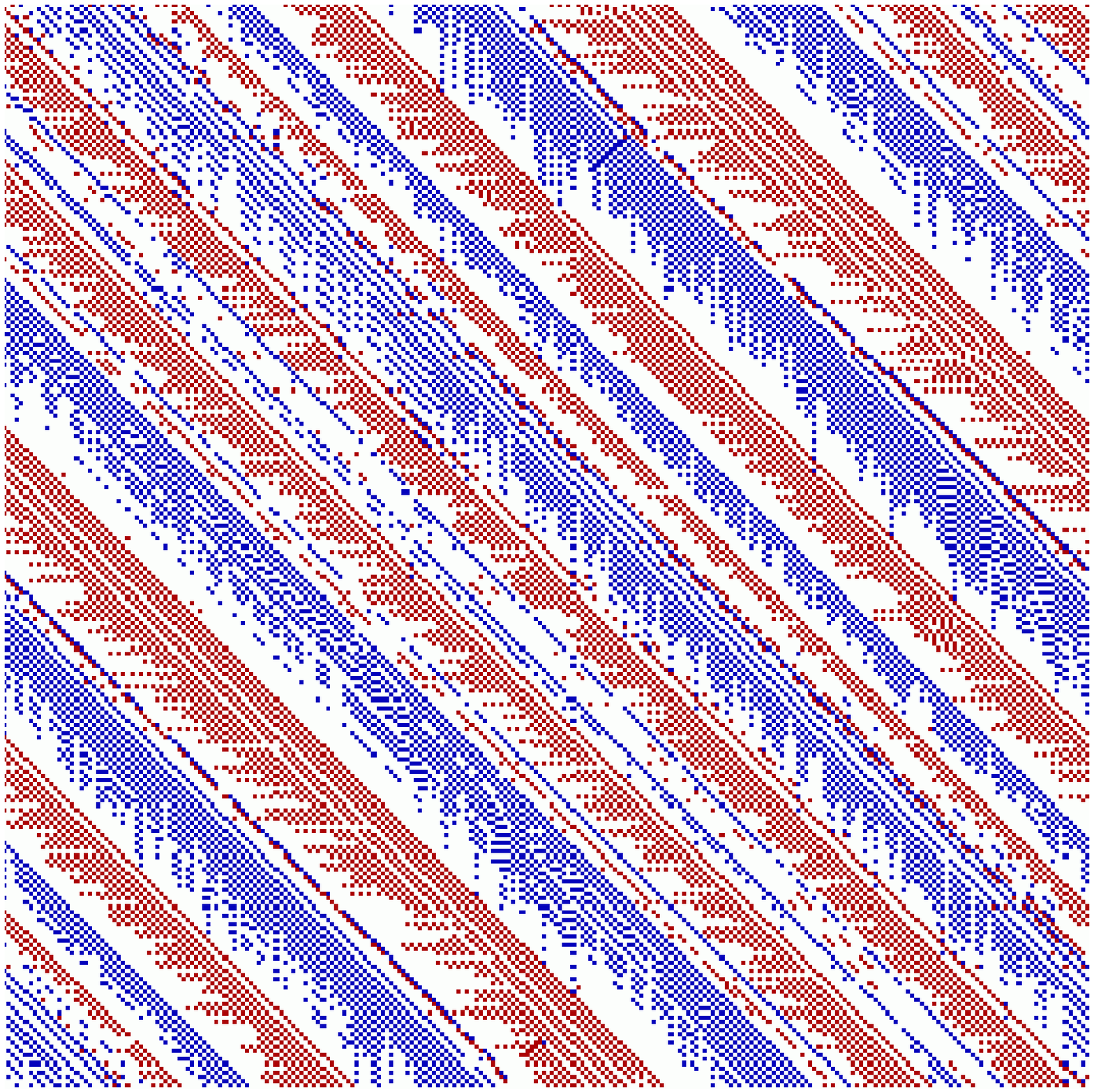}}\hfill\hfill
\resizebox{2in}{!}{\includegraphics{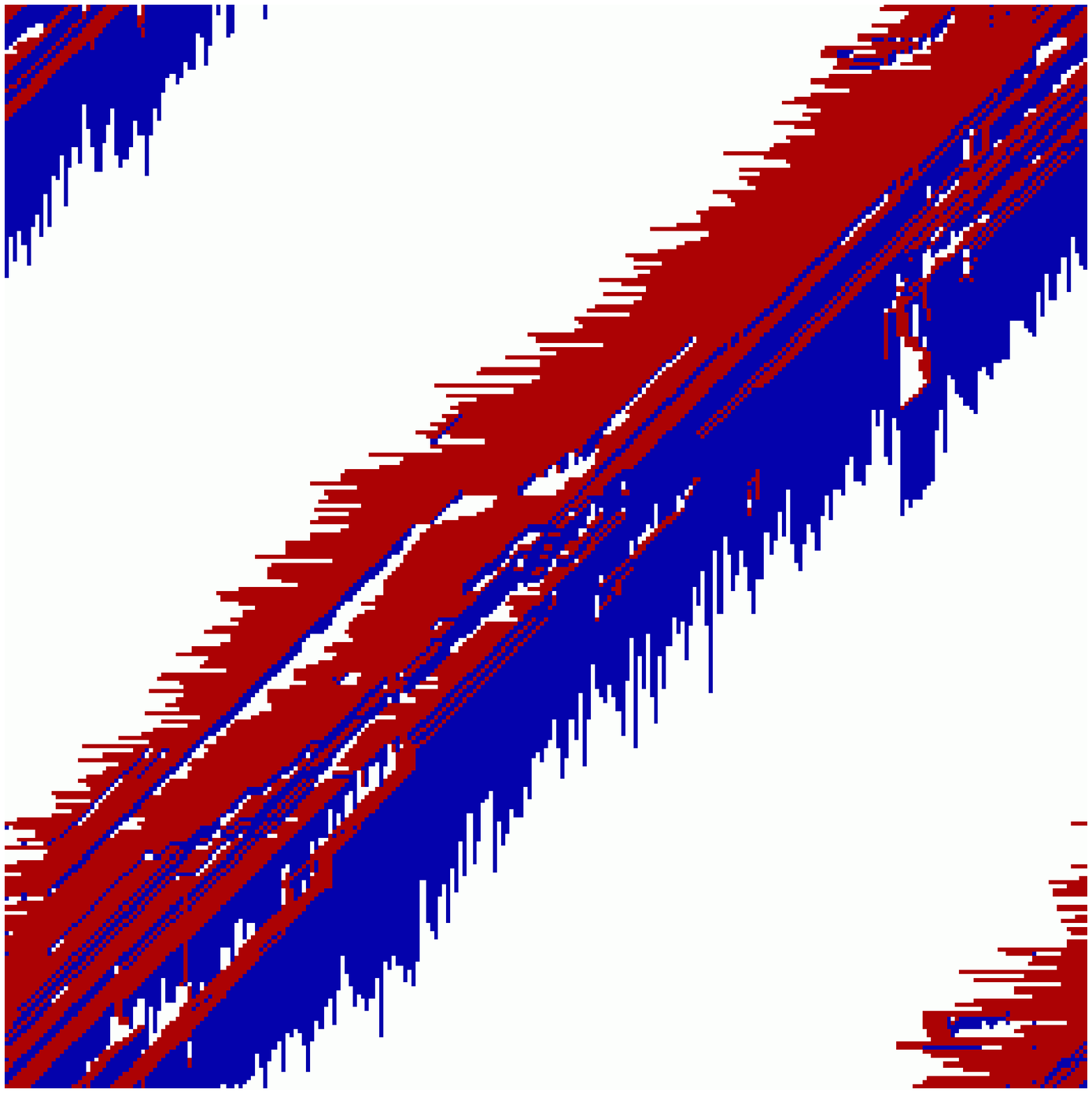}}\hfill\hfill
\resizebox{2in}{!}{\includegraphics{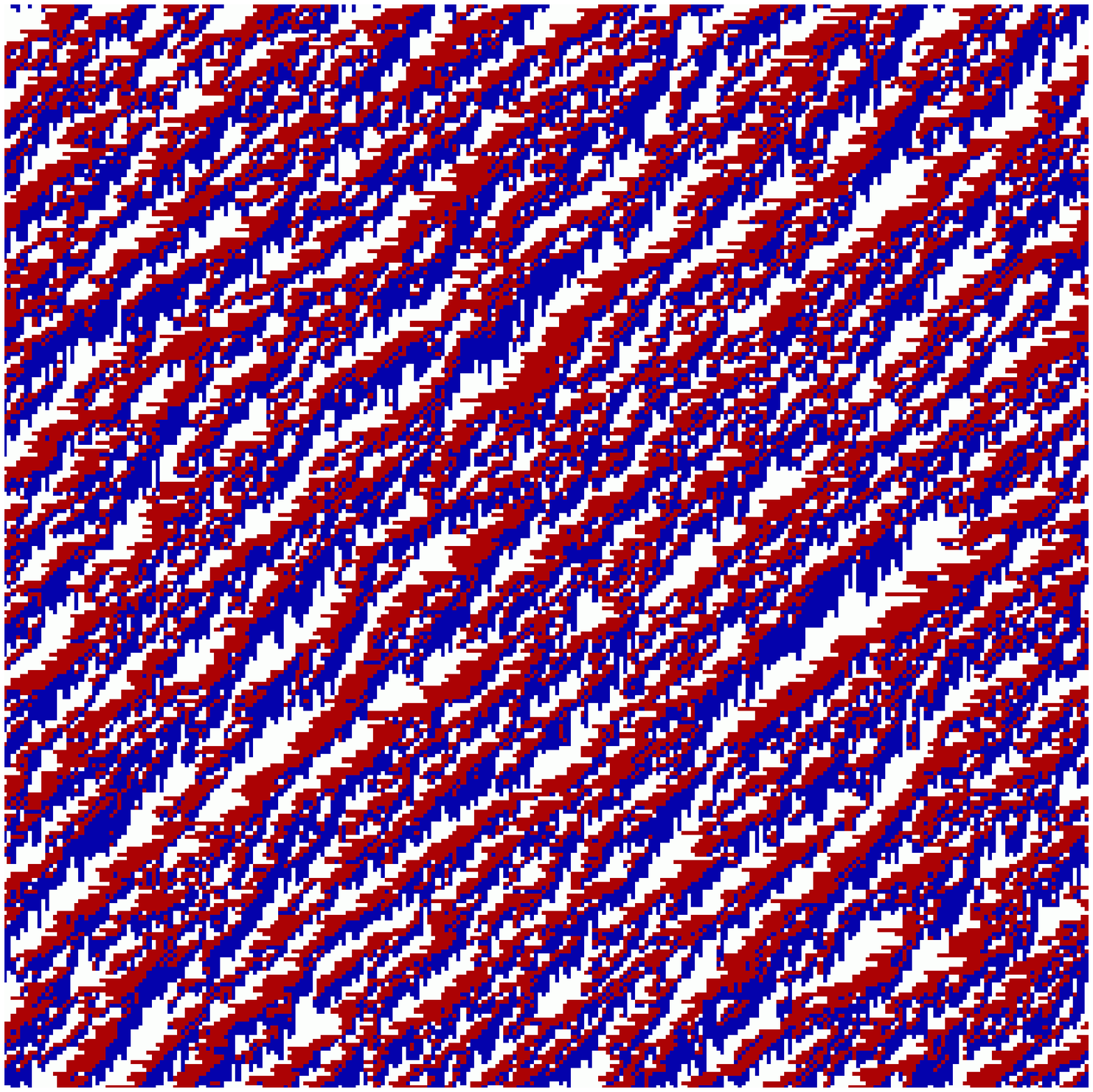}}\hfill}
%\vspace{-0.25in}
\caption{Typical configurations observed for the BML model on an
$L\times L$ system of size $L=256$. (a) The free flowing stage, where
all particles advance during each update ($v=1$).  Note the ordered
stripes of alternating east and north-bound cars.  The width of the
stripes increase, on average, with density $\rho$ while in the low
density phase.  (b) A fully jammed configuration, consisting of one
global jam.  Note the jam length, $\sqrt{2} L$, is larger than the
system size. (c) A high density, random jam configuration.}
\label{fig:final}
\end{figure}

If initialized with a low enough density of cars, the system
eventually self-organizes into a configuration where all cars can move
at each time step (each car has asymptotic velocity equal to unity).
A typical such configuration is shown in Fig.~\ref{fig:final}(a). If
initialized at slightly higher density, the cars are blocked by other
cars, until eventually all cars end up participating in one large
global jam, where no car can move (asymptotic velocity equal to zero).
A typical global jam is shown in Fig.~\ref{fig:final}(b). The
transition between the two behaviors has been thought of as sharp,
showing characteristics of a first order phase transition.
Initialized at much higher densities, small jams begin simultaneously
throughout the lattice and merge almost immediately with other small
jams, leaving all cars blocked (with all velocities equal to zero).
In this high density phase, the system has no time to self organize,
and instead of one global jam, we observe a collection of small random
jams.  This latter type of jam has been compared to traffic in a large
city during ``rush hour'': a car might escape one jam, only to quickly
join the tail of the next.  We expect that for an infinite size
system, the fully jammed state resembles this random type of jam.

Most of the understanding of the BML model has come from numerical
simulations\cite{BML92,Naga93-anisotropy,FukuiIshi93,TadaKiku94,TadaKiku95,greenwave96,MartCuesta95,ChauWanYan98}.
Theoretical analysis has been limited to mean-field
approaches\cite{Naga93-anisotropy,IshiFukui94,Cuesta-LB-95,WangWooHui96},
and attempts to start with continuum hydrodynamic equations and
formulate an equivalent discrete model\cite{Naga99}. There are general
beliefs about this model, that the transition is first order and that
the critical density, $\rho_c$, decreases with increasing system size,
possibly reaching the value $\rho_c=0$ as the system size approaches
infinite\cite{BML92}. The BML model has been simulated extensively,
but there are inconsistencies in the literature and lack of detail of
numerical implementations (such as the size of the ensemble being
averaged over).  Details of numerical studies have been published only
for small systems, on the order of
$L=10-50$\cite{FukuiIshi93,Naga93-anisotropy}. Larger systems have
only been studied coarsely, or in the context of self-organized versus
random jams\cite{TadaKiku94,TadaKiku95}.

Despite extensive numerical simulation, the existence of stable
intermediate phases (with $0 < v < 1$) has not been explicitly
reported previously.  Fukui and Ishibashi\cite{FukuiIshi93} do show
evidence of the existence of an intermediate phase in one plot. They
note that for intermediate values of $\rho$, $v$ ``fluctuates around a
certain value for a long time''. The value shown in their plot is
extremely close to the values we observe (plotted in
Sec.~\ref{sec:results}). Aside from this comment, they do not pursue
the issue further. A careful study by T\"{o}r\"{o}k and
Kert\'{e}sz\cite{greenwave96} contains precise details of their
numerical simulations.  They are studying a variant of BML with faster
convergence times (called the green wave model).  Since it is not
possible to theoretically predict the convergence time, they estimate
it, and apply the following, very reasonable, empirical heuristic.  If
a realization has not reached a state, with $v\approx 0$ or $v \approx
1$, within an allotted time (taken to be 5 times the estimated
convergence time), that realization is discarded.  We can only assume
some of the studies of the BML model may have used a similar criteria
of discarding ``non-converged'' states.  Note that for continuum
models, intermediate phases of jammed wavefronts moving through
otherwise freely flowing traffic have been reported. See for instance
Ref.~\cite{KimaSasa95}.

Sensitivity of the BML model to boundary conditions has been reported
previously.  Mart\'{i}nez, {\em et al.}\cite{MartCuesta95} study the
dilute limit ($\rho \rightarrow 0$), and show that different results
are obtained for an ``entangled'' torus versus a conventional one.
They raise interesting questions about how to get at the bulk
properties using only finite size simulations, but do not quantify nor
pursue the effects further.  Chau, Wan, and Yan\cite{ChauWanYan98}
study the BML model on the torus with random boundary conditions (BCs)
(meaning, particles moving off the right (top) edge reappear at some
randomly selected site on the left (bottom) edge).  They claim $v>0$
whenever $\rho<1$, and hence dismiss such systems as being ``not very
interesting''.  They also note that the velocity and critical density
depend sensitively on the choice of BCs, but they also do not pursue
the effects further.

\section{Simulation Results}\label{sec:results}
We implement the BML model on square lattices of finite size $L\times
L'$, for a range of sizes and varying aspect ratios.  For square
aspect ratios ({\em i.e.}, $L=L'$), the lengths studied range from
$L=64$ to $L=512$. We also study rectangular aspect ratios where the
width, $L$, is an integer multiple of the height, $L'$, and where $L$
and $L'$ are relatively prime.  On each lattice, we implement a range
of densities $\rho$, studying at least ten realizations for each
density.  All simulations are implemented on a special purpose
cellular automata machine, the CAM8\cite{cam8-waterloo}.  The CAM8
performs approximately one-billion site updates per second, comparable
to a modern high-performance desktop computer.  The main advantage of
the CAM8 is that it allows for excellent visualization of the system,
with no overhead in the rendering, giving us live video output of the
dynamics of the system.  As discussed in Sec.~\ref{sec:geom},
visualizing the kinetic pathway gives crucial insight into the
formation of the intermediate states.  

The main subtlety involved with simulating this model is determining
the convergence time ({\em i.e.}, the time is takes to reach the
asymptotic behavior).  All realizations were simulated until converged
($v=0$, $v=1$, or the periodic limit cycle was reached) or for times
out to at least $t=2\times 10^6$ time steps.  Of the realizations
that had not yet converged, many were simulated for orders of
magnitude beyond.  We find this a reasonable compromise, since the
compute power to simulate all samples to times greater than $10^8$ is
beyond our current capacity.  Throughout the remainder of the
manuscript, we refer interchangeably to the east-bound cars as ``red''
and the north-bound as ``blue''.

\begin{figure}[t]
{\hfill(a)\resizebox{2in}{!}{\includegraphics{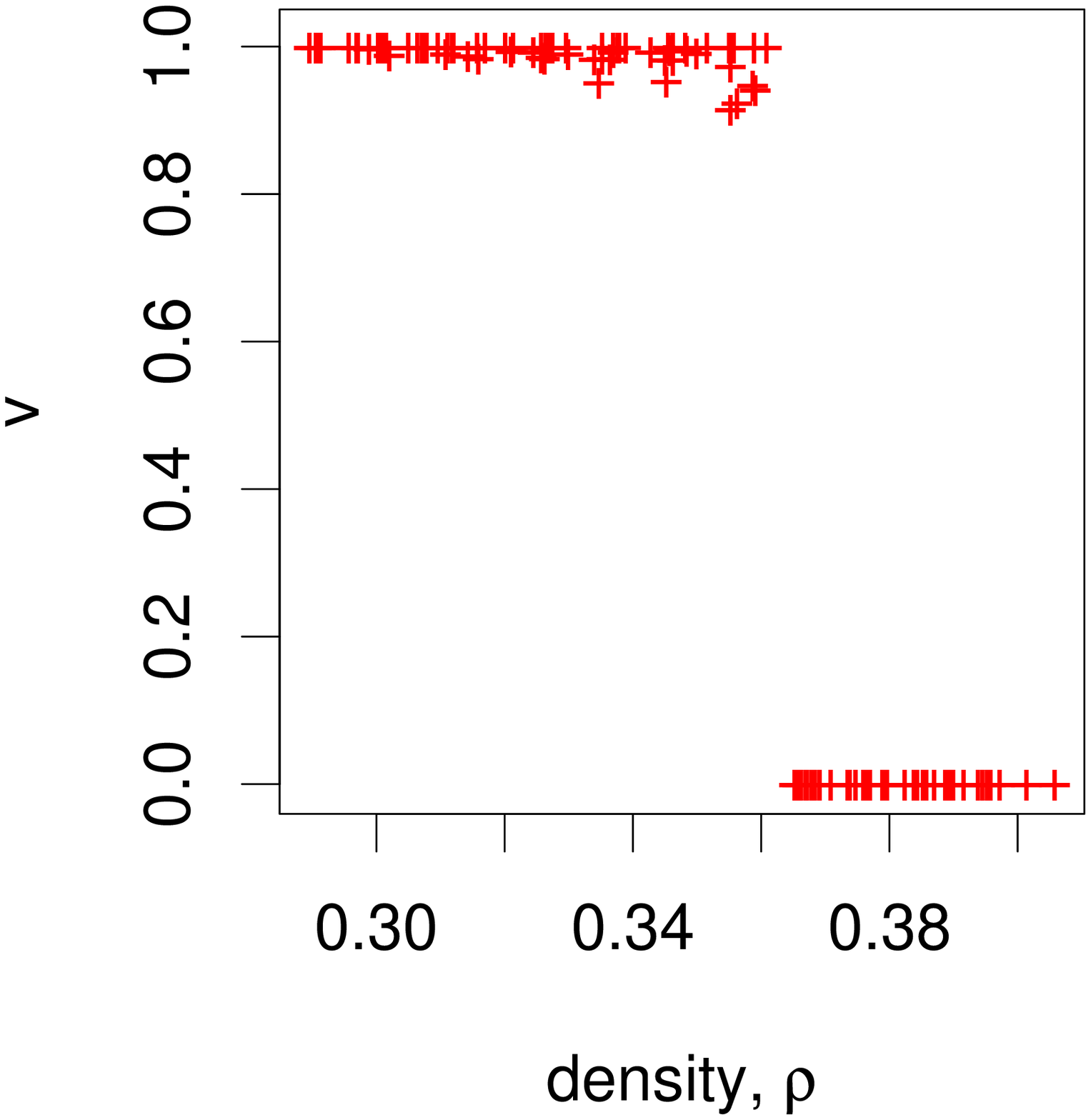}}
\hfill(b)\resizebox{2in}{!}{\includegraphics{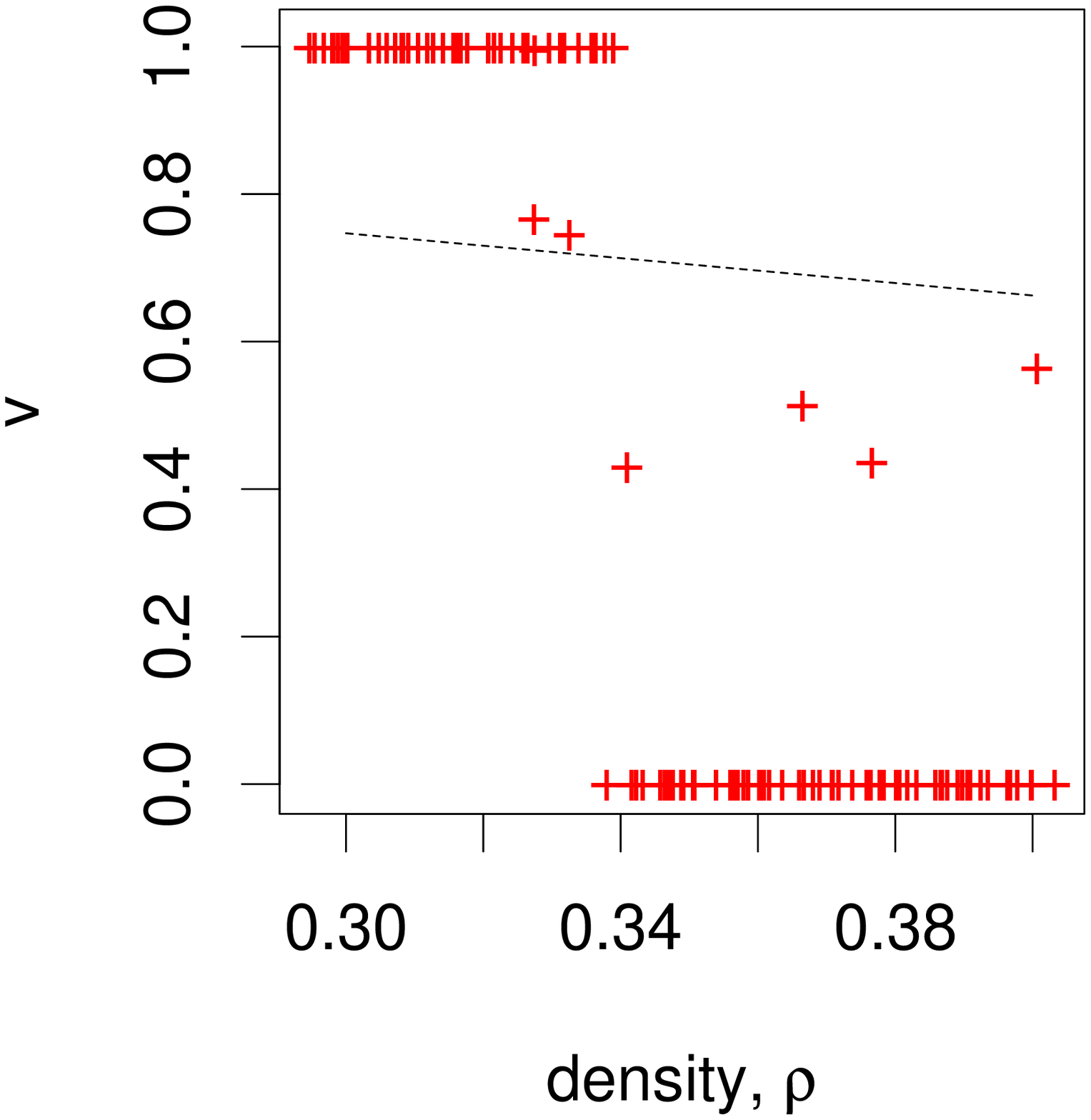}}\hfill\hfill\hfill\\
\hfill(c)\resizebox{2in}{!}{\includegraphics{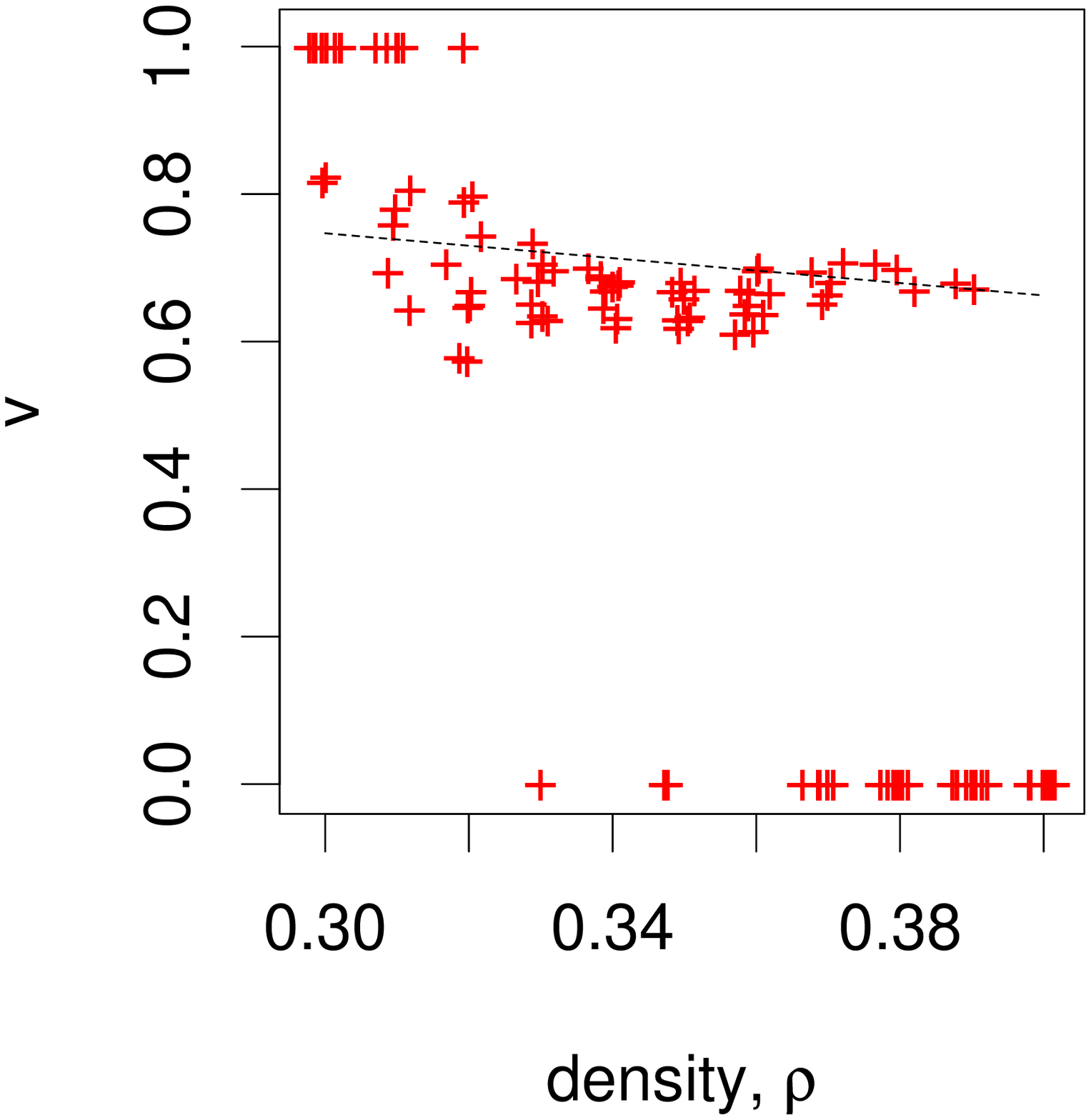}}
\hfill(d)\resizebox{2in}{!}{\includegraphics{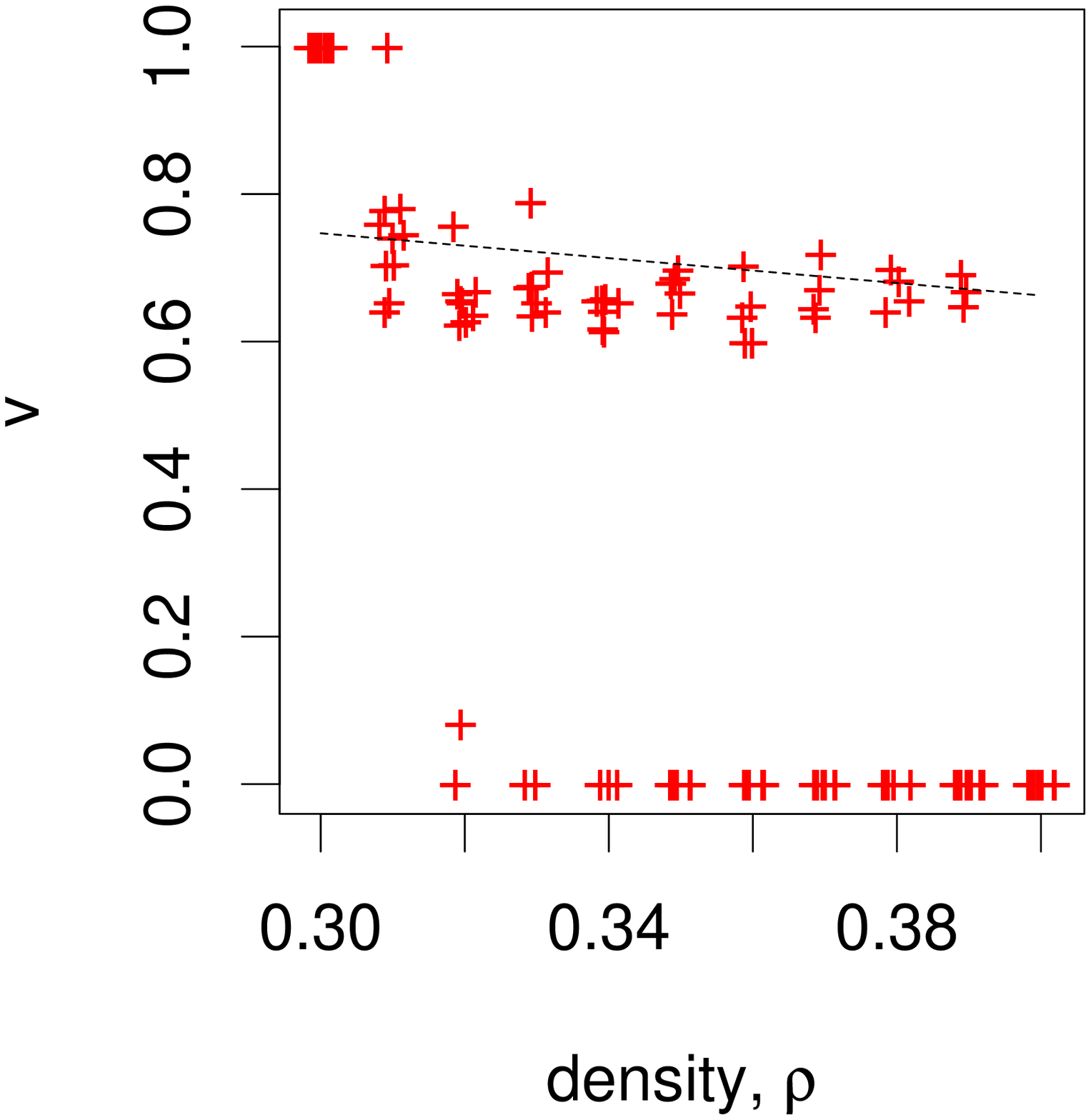}}\hfill}
%\vspace{-0.25in}
\caption{Plot of the average velocity for each individual realization,
$v$, versus the density, $\rho$, for that realization, for an $L\times
L$ lattice: (a) $L=64$, (b) $L=128$, (c) $L=256$, (d) $L=512$.  Note
the emergence of an intermediate phase for $L>64$. The value of $v$
for the intermediate state becomes more crisply defined with
increasing system size, and that the window of coexistence between the
phases broadens. For the $L=512$ system, the average value of $v$ in
the intermediate phase is $\left<v_s\right>=0.673 \pm 0.005$. The
dotted line shown in (b)-(d) is the prediction from Eq.(\ref{eqn:v_s}).}
\label{fig:bml-64-512}
\end{figure}

\subsection{Square aspect ratios}
\begin{figure}[tb]
{\hfill(a)\resizebox{2.25in}{!}{\includegraphics{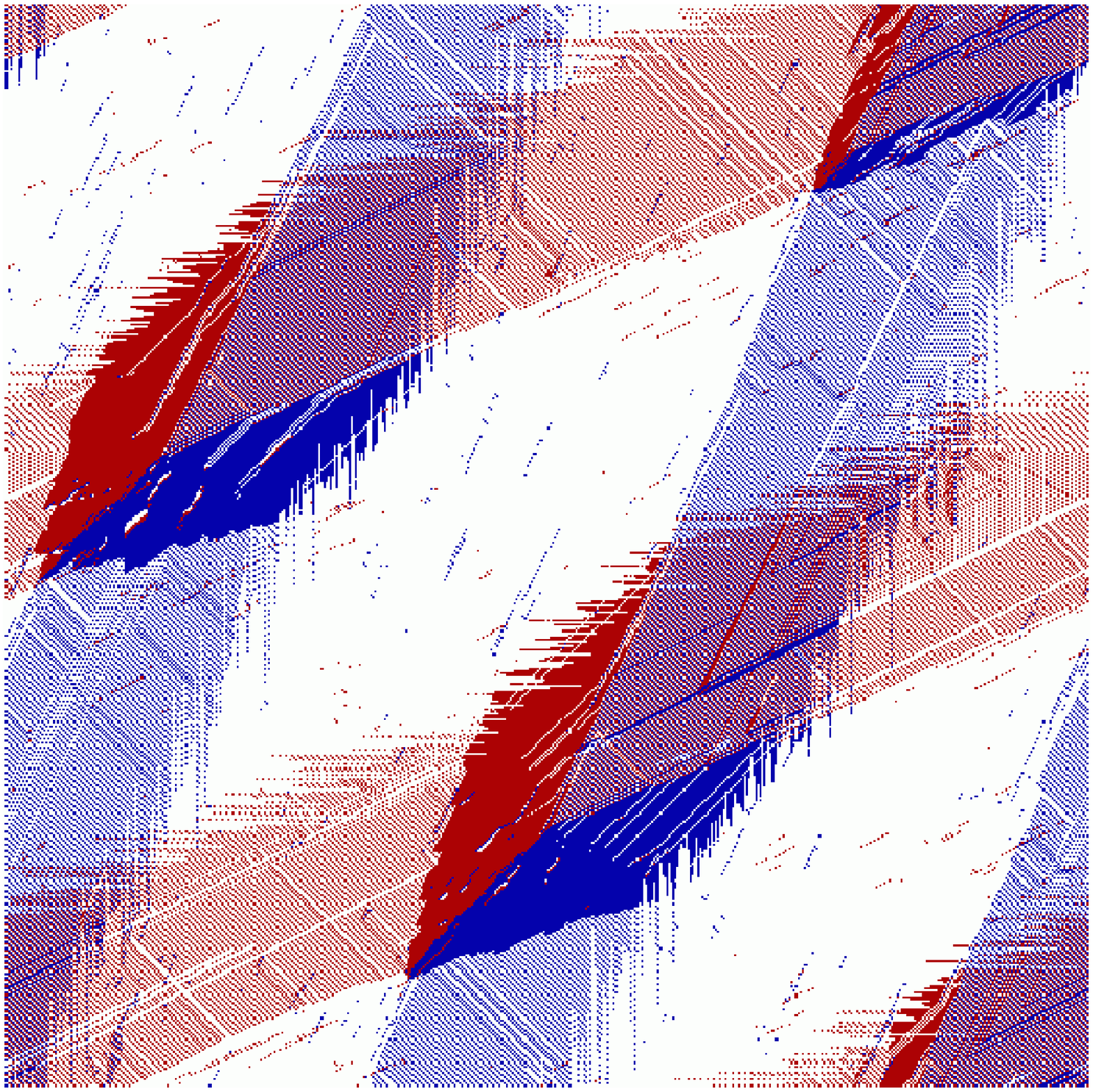}}\hfill\hfill
(b)\resizebox{4.15in}{!}{\includegraphics{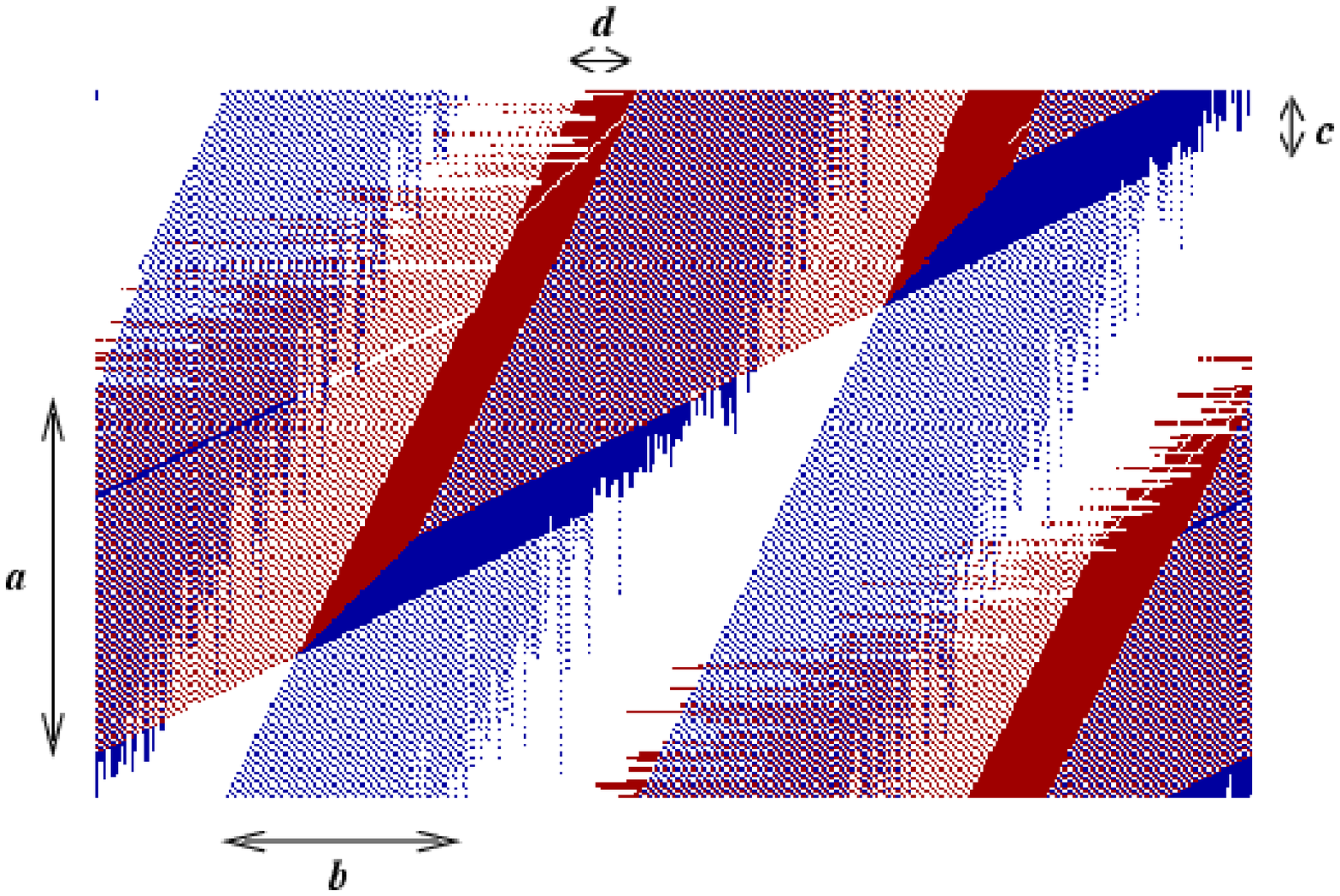}}\hfill}
%\vspace{-0.25in}
\caption{Examples of typical intermediate geometry.  (a) A system
with square aspect ratio where $L=512$, $\wr=1/2$ and $\wb=2$. (b) A
system with rectangular aspect ratio where $L=377$, $L'=233$, $\wr=1$,
and $\wb=3$. Note that in (a) there are many disordered, random cars
in the space between the bands, yet in (b) all cars are ordered.  We
find this crisp order shown by the latter example for all realizations
studied on rectangular aspect ratios with $L$ and $L'$ relatively
prime. If $L$ and $L'$ are not relatively prime, random disordered
cars located between the bands persist. Note that
Fig.(\ref{fig:inter-closeup}) is a close up of the region that has
just shed from the jams in (b).}
\label{fig:intermediate}
\end{figure}

For small size systems we actually observe the predicted behavior of a
sharp transition from freely flowing to total gridlock.
Figure~(\ref{fig:bml-64-512}a) is for an $L \times L$ system with
$L=64$.  It is a plot of the final average velocity observed for each
realization, $v$, versus the density for that particular realization.
Note we are plotting the average velocity for each individual
realization, not an average over all realizations initialized with the
same $\rho$ (hence error bars are on the order of the size of the
plotting symbol used).  Surprisingly, when we implement systems with
$L>64$, we observe a bifurcation where two phases start to coexist, as
we go from low to intermediate values of $\rho$.  The second phase
that emerges has average velocity $v\sim 2/3$, as shown in
Fig.(\ref{fig:bml-64-512}).  In addition, these intermediate states
have a very well defined geometry, of bands of red stripes with slope
one-half, criss-crossing bands of blue stripes with slope two. An
example of the geometry is shown in Fig.(\ref{fig:intermediate}).
Jammed wavefronts are located at the intersections of the bands, and,
as the systems evolves in time, move as solid structures uniformly
down towards the southwest with unit velocity.  Particles are freed
from the head of each jam, but a like number of new jammed particles
aggregate at the tail.  As discussed in detail below, the underlying
lattice imposes constraints on the allowed topologies of the
configurations.  In Sec.~\ref{sec:geom}, we derive a simple
formulation of the constraints.

\begin{figure}[tb]
\rotatebox{270}{\resizebox{3in}{!}{\includegraphics{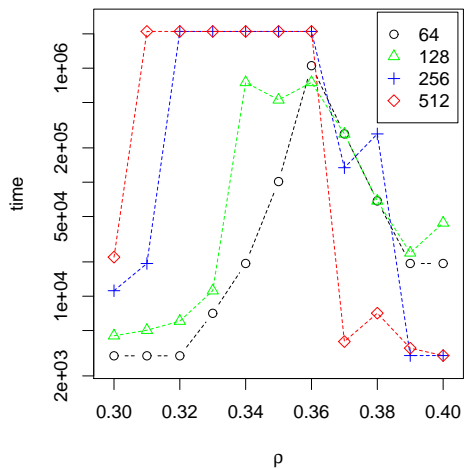}}}
\vspace{-0.1in}
\caption{Median convergence times for the different square aspect ratios
simulated. These results are consistent with previous studies. Note a
value of $t=2\times 10^6$ should be interpreted as $t>2\times 10^6$.}
\label{fig:converge}
\end{figure}

All realizations contributing to Fig.(\ref{fig:bml-64-512}) were
simulated to at least $t=2\times 10^6$ full updates of the entire
space, with various realizations simulated for $t>10^8$.  We do
occasionally observe a realization persist in the $v\sim 2/3$ state
for orders of magnitude, then suddenly jump to either $v=0$ or $v=1$,
with the latter being more common. Regardless, the intermediate state
is at least metastable, persisting for longer than we could simulate
most realizations.  Furthermore it is ``universal'' in the sense that
the value of $v\sim 2/3$ is independent of system size and density,
and all systems that do not go to $v=0$ or $v=1$ go to the same
intermediate state ({\em i.e.}, same geometric structure and
approximate value of $v$). The reason we do not average over the
individual realizations, is that it would obscure the
behavior. Instead of displaying the three distinct quantized states,
averaging would produce a deceitfully smoothly decaying curve.

From our data, it is difficult to determine the exact bifurcation
point where the phases begin to coexist, and the point where they
cease to.  We attempted to identify the factors that distinguished
realizations which converge to $v=0$ from those, with the same
density, that converge to $v\sim 2/3$.  We first investigated
connections to anisotropy, such as an imbalance between the total
number of red versus blue particles.  But we found no correlation.
The probability a realization would jam or go to the intermediate
state is independent of this asymmetry. We also looked at a more
fine-grained measure: the line density of red particles versus the
blue.  Again we found no correlation between this asymmetry and the
likelihood of jamming.

In Sec.~\ref{sec:geom} we discuss the kinetic mechanism observed,
which gives rise to the interleaved band structure exhibited by the
intermediate states.  As mentioned above, the jam interface moves
ballistically, with unit speed, towards the southwest.  The width of
the jam interface can fluctuate.  It seems for the $L=64$ system, the
fluctuations are large enough that eventually the head of one jam
meets the tail of the previous, continuing until eventually one global
jam forms.  Kinetic mechanisms aside, it could be that the convergence
times for $L=64$ are short enough that the intermediate state cannot
be considered metastable.  Fig.(\ref{fig:converge}) shows the median
convergence times observed for the systems with square aspect ratios.
Note that the a value of $t=2\times 10^{6}$ really means $t>2\times
10^{6}$.

\subsection{Rectangular aspect ratios}
\begin{figure}[tb]
{\hfill\resizebox{2.5in}{!}{\includegraphics{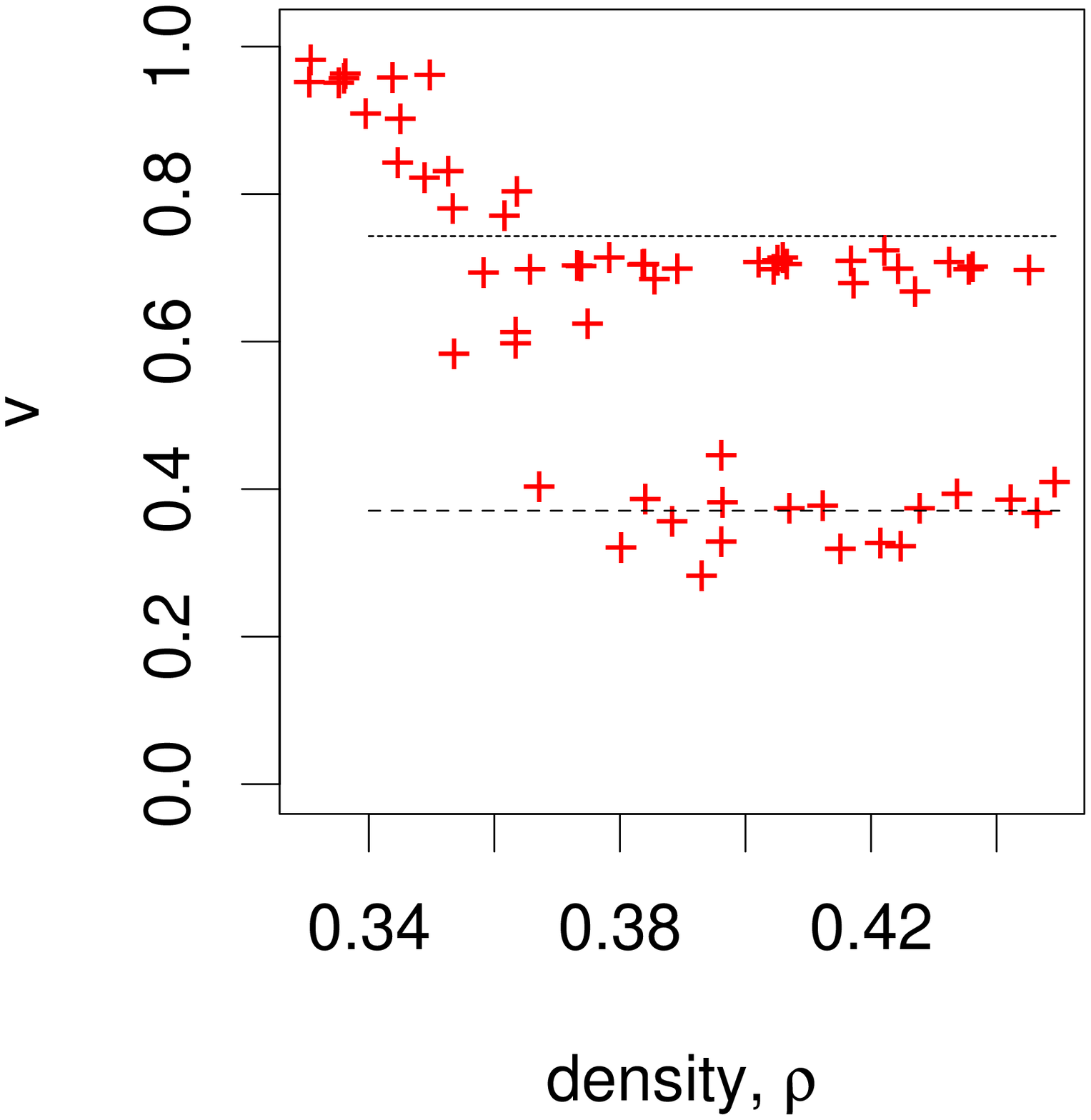}}\hfill\hfill
\resizebox{2.5in}{!}{\includegraphics{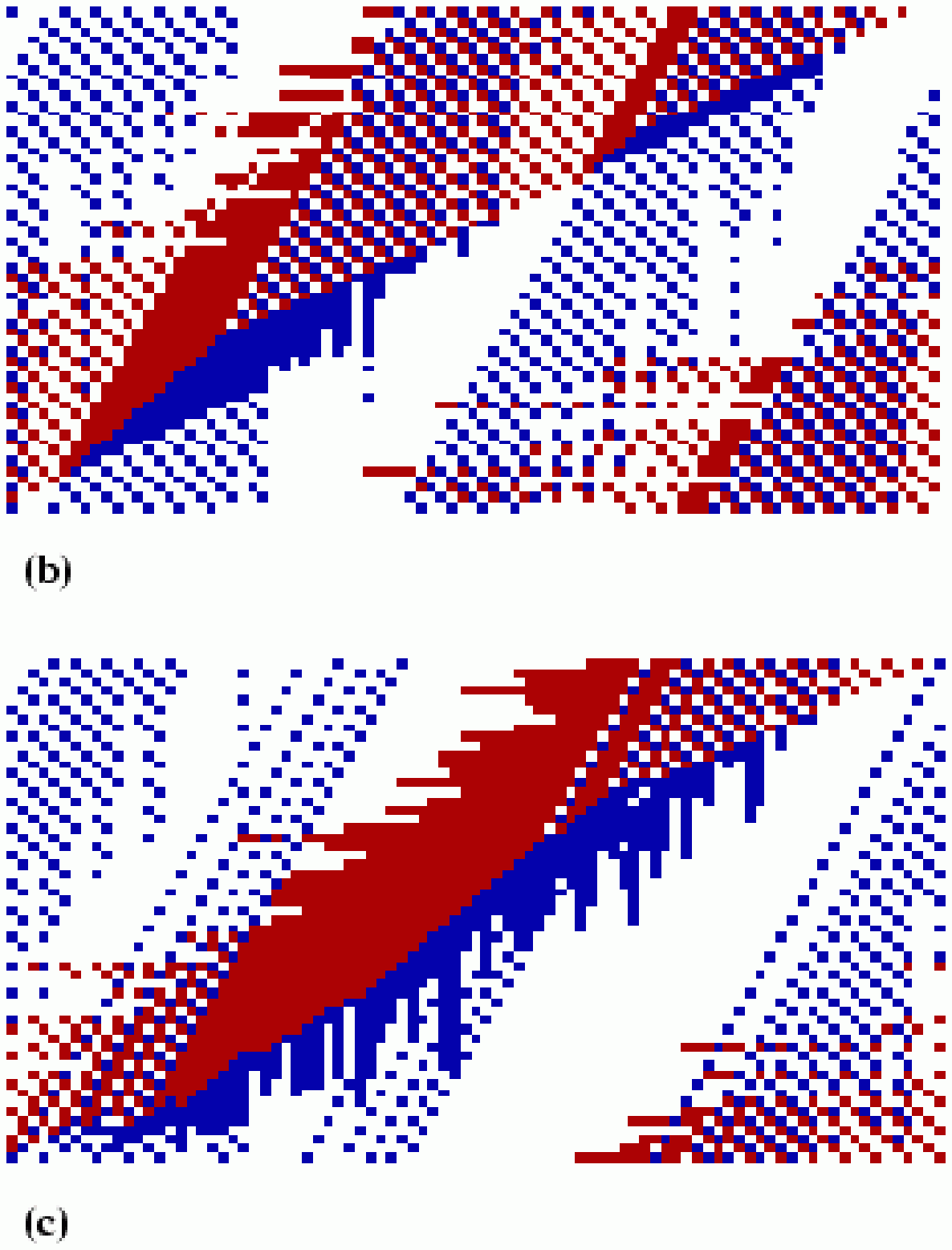}}\hfill}
%\vspace{-0.25in}
\caption{(a) The average velocity for each individual realization,
$v$, versus the density, $\rho$, for that realization, for an $L\times
L'$ lattice, with $L=89$ and $L'=55$. Note the appearance of yet
another well defined intermediate state. Note also, the bifurcation
point, where phase-coexistence ceases, has not yet been reached,
despite the range of $\rho>0.44$.  The dotted lines are the predicted
velocities from Eqs.(\ref{eqn:vr1}) and (\ref{eqn:vr2}),
$v_{r1}\approx 0.7430$ and $v_{r2}\approx 0.3707$. The empirically
determined average values are respectively, $\left<v_{r1}\right> =
0.700 \pm 0.002$ and $\left<v_{r2}\right> = 0.364 \pm 0.004$.  The
geometries of the two types of intermediate states are distinct: (b) A
typical configuration with $v\sim 2/3$. Interface slope $s=1$; (c) A
typical configuration with $v\sim 2/5$. Interface slope $s\approx
2/3$.}
\label{fig:fibo-55x89}
\end{figure}

We also implement the BML model on systems with varying rectangular
aspect ratios.  In particular we study lattices where the two lattice
lengths are subsequent Fibonacci numbers.
Figure~(\ref{fig:fibo-55x89}a) is the plot analogous to those shown in
Fig.(\ref{fig:bml-64-512}), for a system with $L=89$ and $L'=55$. Note
we see the same intermediate velocity of $v\sim 2/3$, but we also see
the emergence of one more possible phase with $v\sim 2/5$.  All the
configurations with $v \sim 2/3$ resemble the one shown in
Fig.(\ref{fig:fibo-55x89}b), with one red band wrapping around the
$\hat{x}$-axis, and three blue bands.  The jam points are at the
intersections of the bands.  Note the crisp, regular geometry.
Figure~(\ref{fig:intermediate}b) is for an equivalent, but larger,
system with the same Fibonnaci aspect ratio but with $L=377$ and
$L'=233$. This figure more clearly illustrates the highly ordered
geometry.  It also includes the definitions of several of the
parameters used in the analysis in the subsequent sections.  All
realizations with $v\sim 2/5$ resemble the one shown in
Fig.(\ref{fig:fibo-55x89}c), with one red band and approximately two
blue bands (though the latter are not so clearly defined).

\begin{figure}[b]
{\hfill(a)\rotatebox{270}{\resizebox{1.75in}{!}{\includegraphics{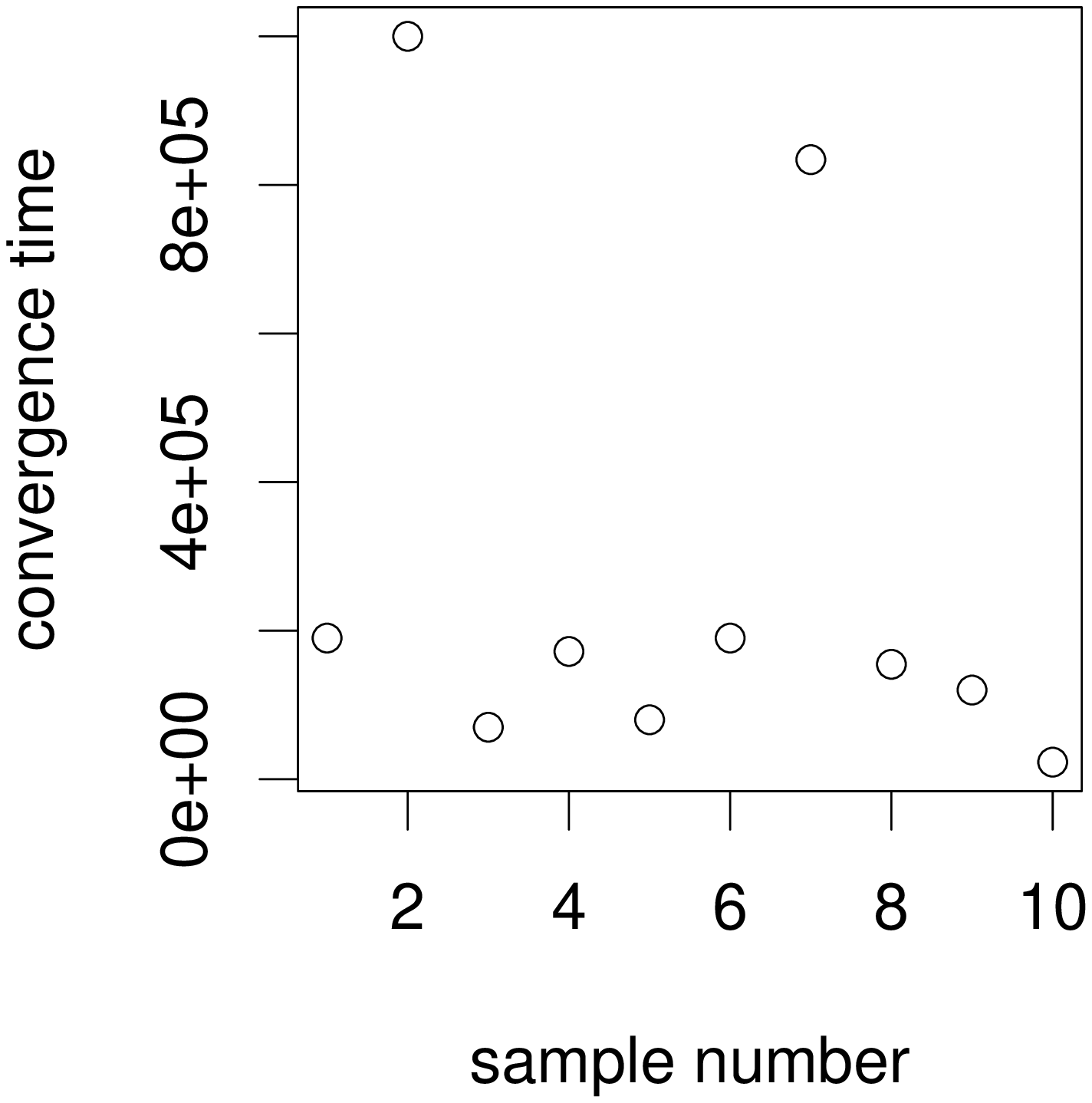}}}\hfill\hfill(b)\rotatebox{270}{\resizebox{1.75in}{!}{\includegraphics{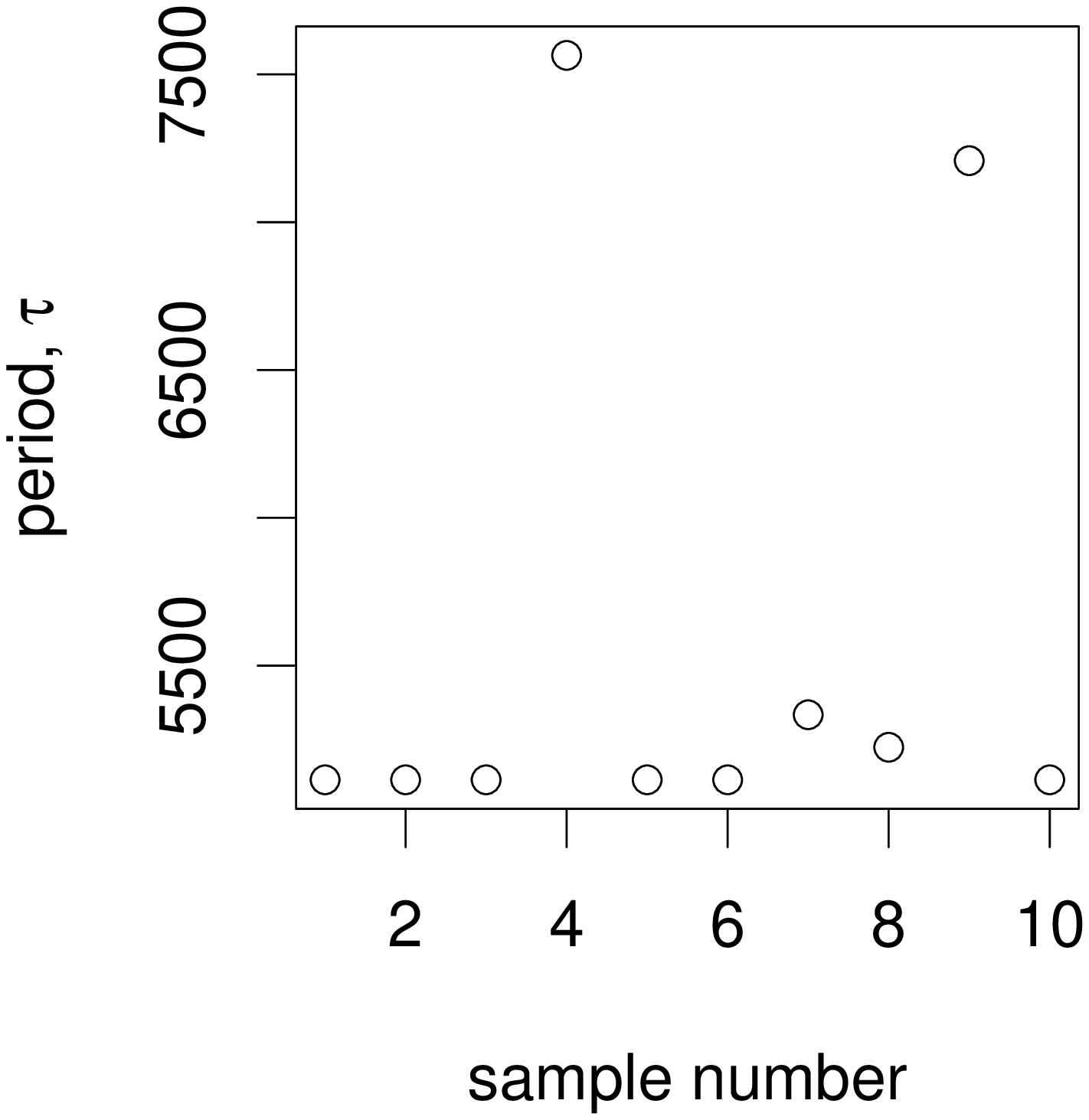}}}\hfill}
%\vspace{-0.25in}
\caption{For simulations with $L=89$ and $L'=55$ and $\rho=0.38$, we
plot the sample number versus, (a) the time to reach the periodic
limit cycle; (b) the period of the cycle, $\tau$.  Eight of the ten
realizations simulated reach final configurations of type I, as shown
in Fig.(\ref{fig:fibo-55x89}b). They all have the shorter values of
$\tau$. Surprisingly six of these eight realizations have the same
period, $\tau=5114$, though their microscopic configurations differ.
The other two realizations, with significantly larger values of
$\tau$, are of type II, as shown in Fig.(\ref{fig:fibo-55x89}c).}
\label{fig:convTau}
\end{figure}

One of the most striking differences when comparing these rectangular
aspect ratios to the square, is that for the rectangular, the
intermediate configurations are exactly periodic with a short
recurrence time: the exact microscopic configuration of particles
repeats every $\tau$ updates.  We observe systems of size $L=89$ and
$L'=55$, settling into the periodic behavior typically in a time less
than $t \sim 100,000$ updates, with a period on the order of $\tau
\sim 6000$ updates. Figure~\ref{fig:convTau} is a plot illustrating
such typical behavior.  Another striking difference, is that not one
realization has jammed fully ($v=0$), despite the fact that the
largest density simulated is $\rho = 0.45$.  Note that for the square
aspect ratios, the bifurcation point where the phases cease to coexist
is at $\rho \approx 0.40$.  Another striking difference is the lack
of disorder for the relatively prime systems with rectangular aspect
ratios.  In Fig.(\ref{fig:intermediate}a) there are isolated particles
(``dislocations'') moving in the area between the bands.  However for
the rectangular case, all the particles manage to join the ordered
bands.

\section{Geometric considerations}\label{sec:geom}
\subsection{Kinetic pathways}\label{subsec:kinetics}
By watching the dynamical evolution of the system, starting from the
initial configuration, the mechanism by which the intermediate phases
form can be observed.  Often one global jam initially begins to form,
yet the head of the jam just fails to meet up with the tail, leaving a
few lattice sites of distance between them.  Particles shed from the
head as soon as allowed by the local environment (since all particles
move whenever possible), leaving with a well defined order.  See
Ref.~\cite{bml-webpage}, for video images of the dynamical evolution.

\begin{figure}[t]
\resizebox{5in}{!}{\includegraphics{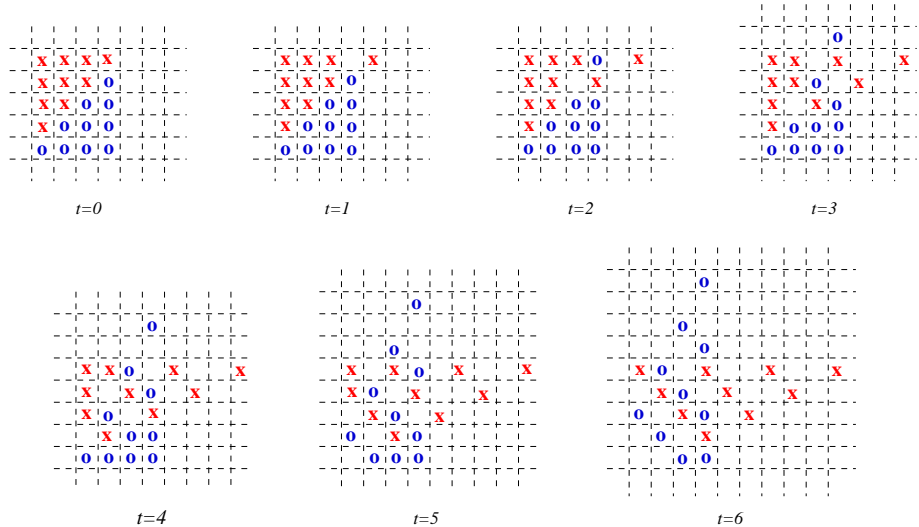}}
\vspace{-0.15in}
\caption{Shedding from a jam.}
\label{fig:shed}
\end{figure}

To understand the pattern formed by the shedding, consider first one
row of a solid isolated block of red particles.  Since the particles
only advance provided the site they wish to occupy is empty, a
particle would leave the head of the jam only every other timestep.
However, we can now consider a diagonal interface formed by a
triangular block of red particles in contact with a triangular block
of blue particles. See Fig.(\ref{fig:shed}). Note red (eastbound)
particles are represented by ``x'' and blue (northbound) particles by
``o''.  Each subsequent time step illustrated corresponds to one
complete update of the space ({\em i.e.}, one north-step followed by
one east-step).  Recall, all particles of the same species update
synchronously.  Step 1: No ``o's'' are able to move. But, during the
east-phase of the timestep, the first ``x'' moves away from the jam,
opening up a space.  Step 2: The first ``o'' advances.  This blocks
all other ``x's'' in the original row, yet opens up a space for an
``x'', one site from the original ``x'' along the southwest diagonal,
to advance.  The original ``x'' to move, also continues advancing.
Step 3: That first ``o'' continues advancing and opens up a space for
an ``x'' in the original row to advance---two time steps delayed from
the first ``x'' to move in that row.  The ``o'' below this original
``o'' is currently blocked. However an ``o'' one site from the
original along the southwest diagonal is now free to advance. Note
this ``o'' blocks the ``x'' in its newly occupied row from advancing,
yet opens up a space for the ``x'' one row southward to advance.
Hence a pattern emerges: a red particle sheds from within the same row
only every third timestep, yet from a site further southwest every
timestep, yielding red bands of density $\rho_r = 1/3$ with slope
$s_r=1/2$.  Likewise for the blue particles, a blue particle sheds
from within the same column only every third timestep, yet from a
successive site along the southwest every timestep, yielding blue
bands of density $\rho_b = 1/3$ with slope $s_b = 2$.  The jams occur
at the intersections of the bands, and the interface of a jam has
slope $s$. Typically, $s=1$. In Fig.(\ref{fig:shed}) one can see the
order beginning to emerge.

\begin{figure}[t]
\resizebox{3.5in}{!}{\includegraphics{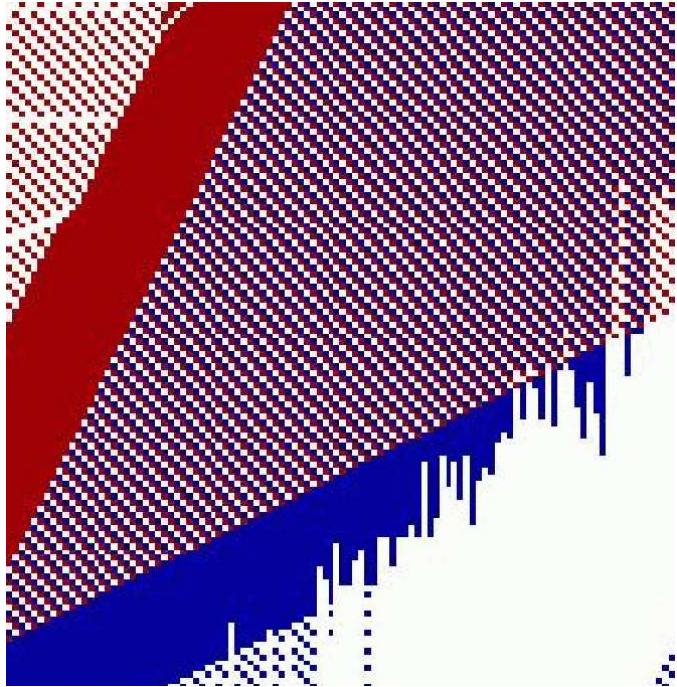}}
\caption{A zoomed in view of a jam shown in Fig.(\ref{fig:intermediate}b).}
\label{fig:inter-closeup}
\end{figure}

More illuminating is to view a closeup near the jammed regions.
Fig.(\ref{fig:inter-closeup}) is a zoomed in view of the region near
one of the jams shown in Fig.(\ref{fig:intermediate}b).  Note the
order that exists in the region above the jam interface, which has
shed from the jam in the manner described above.  In this region there
are alternating diagonal stripes of blue, red, and empty.  The stripes
have ``phase locked''---on the next update the blue stripes move into
the empty stripe regions, leaving room for the red stripes to move on
the subsequent update, and so on.  Hence the system has organized
itself into the highest density packing that still allows all
particles to move with $v=1$.
% say something about non-interacting lattices????!!!!

\subsection{Winding number}\label{subsec:winding}
Since the system lives on a torus, the bands must wrap seamlessly
around it.  Noting these facts, we can develop a mathematical
expression for the number of red bands, $\omega_r$ and the number of
blue, $\omega_b$, (referred to respectively as the ``winding'' number
for red and blue), that must be present.  Essentially, we can
calculate the length of the regions of slope-$1/2$, slope-$s$, and
slope-$2$ that must be present for the bands to wrap around the torus.
(The slopes are respectively those of the red band, the interface, and
the blue band).  Consider an $L\times L'$ lattice, and a red band
starting in the lower left hand corner.  Moving out along the
$\hat{x}$-direction, there are $k$ sites with slope-$1/2$ and $k'$
sites with slope-$s$.  Similarly for a blue band starting in the lower
left hand corner, there are $m$ sites with slope-$2$ and $m'$ sites
with slope-$s$.  The constraints are:
\begin{eqnarray}
k + k' & = & L \label{eqn:k1} \\
\frac{1}{2} k + sk' & = & \omega_r L' \label{eqn:k2} \\
m + m' & = & L \\
2 m + sm' & = & \omega_b L' \\ \label{eqn:m2}
k' & = & m', \label{eqn:interface}
\end{eqnarray}
where, due to the lattice, $\{k, k', m, m' \}$ are positive integers,
and $\{ \omega_r,\omega_b \}$ are either positive integers or equal to
$1/g$, where $g$ is a positive integer ({\em e.g.}, for the
configuration shown in Fig.(\ref{fig:intermediate}a), $\wr=1/2$).  The
final equation, Eq.~(\ref{eqn:interface}), expresses that the length
of the jam interface must be the same for the red and blue cars.
Combining this system (of five equations in seven unknowns), we can
solve some combination of variables in terms of the others. Solving
first for $k'$:
\begin{equation}\label{eqn:kprime}
k' = \left(2\wr L' - L\right)/(2s-1).
\end{equation}
Using this, we can solve for $\omega_b$ in terms of $\omega_r$, $s$,
$L$, and $L'$, and obtain:
\begin{equation}\label{eqn:winding}
\wb = \frac{2L}{L'} - \frac{(2-s)}{(2s-1)}\left(2\wr-\frac{L}{L'}\right).
\end{equation}

\begin{figure}
\resizebox{4.5in}{!}{\includegraphics{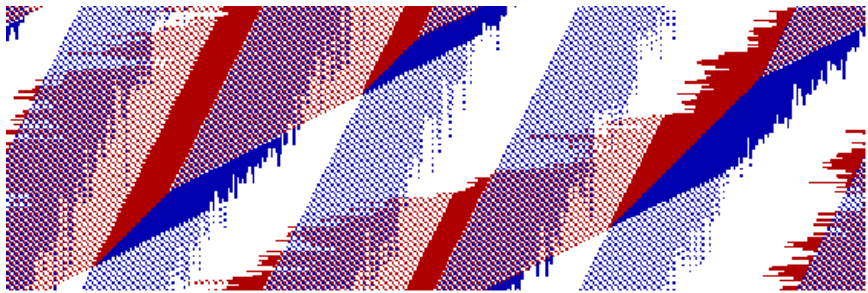}}
\caption{A typical realization with $L=769$ and $L'=256$ ({\em i.e.}. $L =
3L'+1$, where the additional lattice site is to make the lengths
relatively prime).  Note, that as predicted in
Table~\ref{table:winding}, the configuration is consistent with $s=1$,
$\wr = 2$, and $\wb=5$.  Realizations with $L=768$ and $L'=256$ ({\em
i.e.}, $L=3L'$) have a similar structure, but surprisingly lack the
crisp order, and instead have disordered cars at random locations
between the bands.  For images of the latter, see
Ref.~\cite{bml-webpage}.}
\label{fig:bml-aspect3}
\end{figure}

Knowing that the interface slope $s\sim 1$, we can tabulate the
allowed values of $\omega_r$ and $\omega_b$ in terms of the aspect
ratio of the space, $L/L'$. The allowed values for various aspect
ratios are listed in Table~\ref{table:winding}.  Note, the value
$\omega_r=1/2$ means the red band has only reached height $L'/2$ in
traversing distance $L$ (as in Fig.(\ref{fig:intermediate}a)).  We
implement systems with the various aspect ratios shown in
Table~\ref{table:winding}, and find the experimental configurations
observed all match the predicted behavior.  See in addition to the
previous figures, Fig.(\ref{fig:bml-aspect3}).  The final lines in
Table~\ref{table:winding} are for systems where we observe empirically
the values of $\wr$ and $\wb$, and using Eq.(\ref{eqn:winding}) can
predict the value of $s$.  Recall $\wr$ and $\wb$ must be integers.
The system seems to tune the value of $s$ to allows this.  For
instance, if $L$ and $L'$ are successive Fibonacci numbers ({\em
i.e.}, $L/L' = (1+\sqrt{5})/2$), a rearrangement of
Eq.(\ref{eqn:winding}) predicts $s=1.17$, which matches our empirical
observations.  Note in Fig.(\ref{fig:intermediate}b), the upper jam
has small glitches where $s>1$.
\begin{table}
\begin{tabular}{ccccc}
$L/L'$ & \ \ \ \ \ \ $s$ \ \ \ \ & \ \ \ $\omega_r$ \ \
\ & \ \
\ $\omega_b$ \ \ \ & \ \ \ \ \ \ $v $ \ \ \ \ \ \ \\ \hline
1 &  1 &  1/2 & 2 & $0.673 \pm 0.005$ \\
1 &  1 &   1  & 1 & 0 \\
2 &  1 &   1  & 4 & $\sim$ 2/3\\
2 &  1 &   2  & 2 & 0 \\
5/3 & 1 &  1  & 3 & $\sim$ 2/3\\
3 &  1 &   2  & 5 & $\sim$ 2/3\\
{\scriptsize{(1+$\sqrt{5}$)/2}} & 1.17 & 1 & 3 & $0.700 \pm 0.002$\\
{\scriptsize{(1+$\sqrt{5}$)/2}} & 7/10 & 1 & 2 & $0.364 \pm 0.004$\\
\end{tabular}
\caption{Allowed winding numbers, $\wr$ and $\wb$, for lattices with
different aspect ratios, $L/L'$, and interface slope $s$. The values
of $v$ which include error bars are from our numerical simulations.}
\label{table:winding} 
\end{table}

\subsection{Average velocity}
Figure~\ref{fig:intermediate}, illustrates the typical geometry for
realizations with $v\sim 2/3$.  In this example, $\wr=1$ and
$\wb=3$.  We label on the figure the lengths $a$, $b$, $c$, and $d$,
denoting respectively the width of the red band, the blue band, the
blue jam and the red jam.  We define two discontinuous functions which
will simplify notation later on:
\begin{equation}
\Theta_r(\wr) = \left\{ \begin{array}{ll}
                1 & {\rm \ \ if \ \ } \wr \ge 1 \\
                1/\wr & {\rm \ \ if \ \ } \wr < 1,
                \end{array}
                \right.
\end{equation}
\begin{equation}
\Theta_b(\wb) = \left\{ \begin{array}{ll}
                \wb & {\rm \ \ if \ \ } \wb \ge 1 \\
                1 & {\rm \ \ if \ \ } \wb < 1,
                \end{array}
                \right.
\end{equation}
In other words, $\Theta_r$ is the number of independent red bands in
one column of the lattice.  $\Theta_b$ is the number of independent
blue bands in one row of the lattice.  If we denote the density of the
red band as $\rho_a$, we can express the average number of red
particles in a column of the space:
\begin{equation}\label{eqn:a}
\Theta_r(\wr) \rho_a a = \rho L'/2.
\end{equation}
Likewise, the average number of blue particles in a row:
\begin{equation}\label{eqn:b}
\Theta_b(\wb) \rho_b b = \rho L/2.
\end{equation}
Recall $\rho$ is the overall particle density (red plus blue).
Empirically we determined that $\rho_a = \rho_b = 1/3$, which is a
consequence of the dynamics described in Sec.~\ref{subsec:kinetics}.
Using the basic equations described in Sec.~\ref{subsec:winding} and
knowledge of the ``typical'' geometry, we can solve for the velocity
of the intermediate state.  Unfortunately, we have to consider the
square and rectangular aspect ratios independently, and do not have
one equation that describes all cases.  The assumptions described
above are valid for the rectangular aspect ratios, but fail to capture
the full behavior of the square ones.

\subsubsection{Rectangular aspect ratios, type I:} Configurations such as
the one shown in Fig.~(\ref{fig:intermediate}) have a regular, highly
structured, geometry that is well described by the formalism in
Sec.~\ref{subsec:winding}.  We find that the values of $k'$ predicted
by Eq.~\ref{eqn:kprime} exactly match those obtained experimentally.
Typically there are $n_j$ jams, with the particles equally divided
amongst them. The interface width per jam is $k'_i = k'/n_k$, where
$n_k$ is the total number of jams one red band is involved in, as it
wraps once around the $\hat{x}$-axis.  See
Fig.(\ref{fig:intermediate}a), for example, where $n_k\approx 2$, and
$n_j=3$.

\begin{figure}[tb]
\hfill\resizebox{2in}{!}{\includegraphics{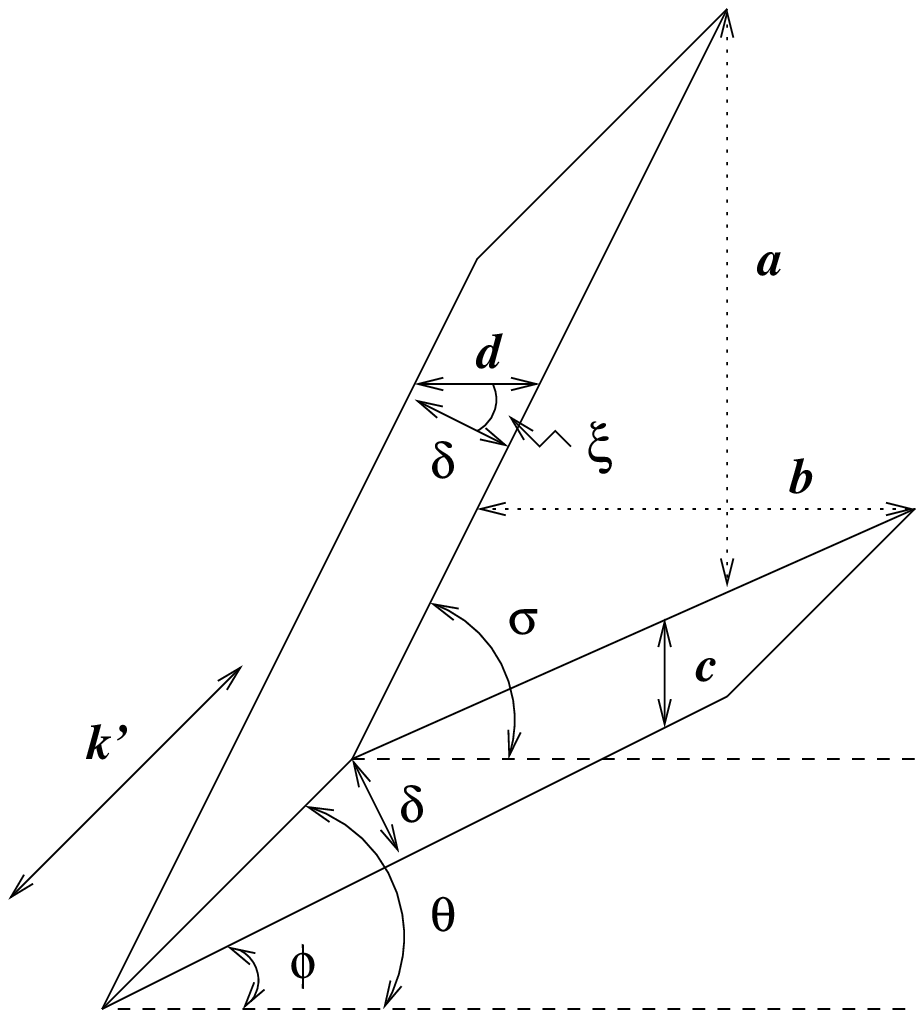}}\hfill\hfill
\resizebox{1.5in}{!}{\includegraphics{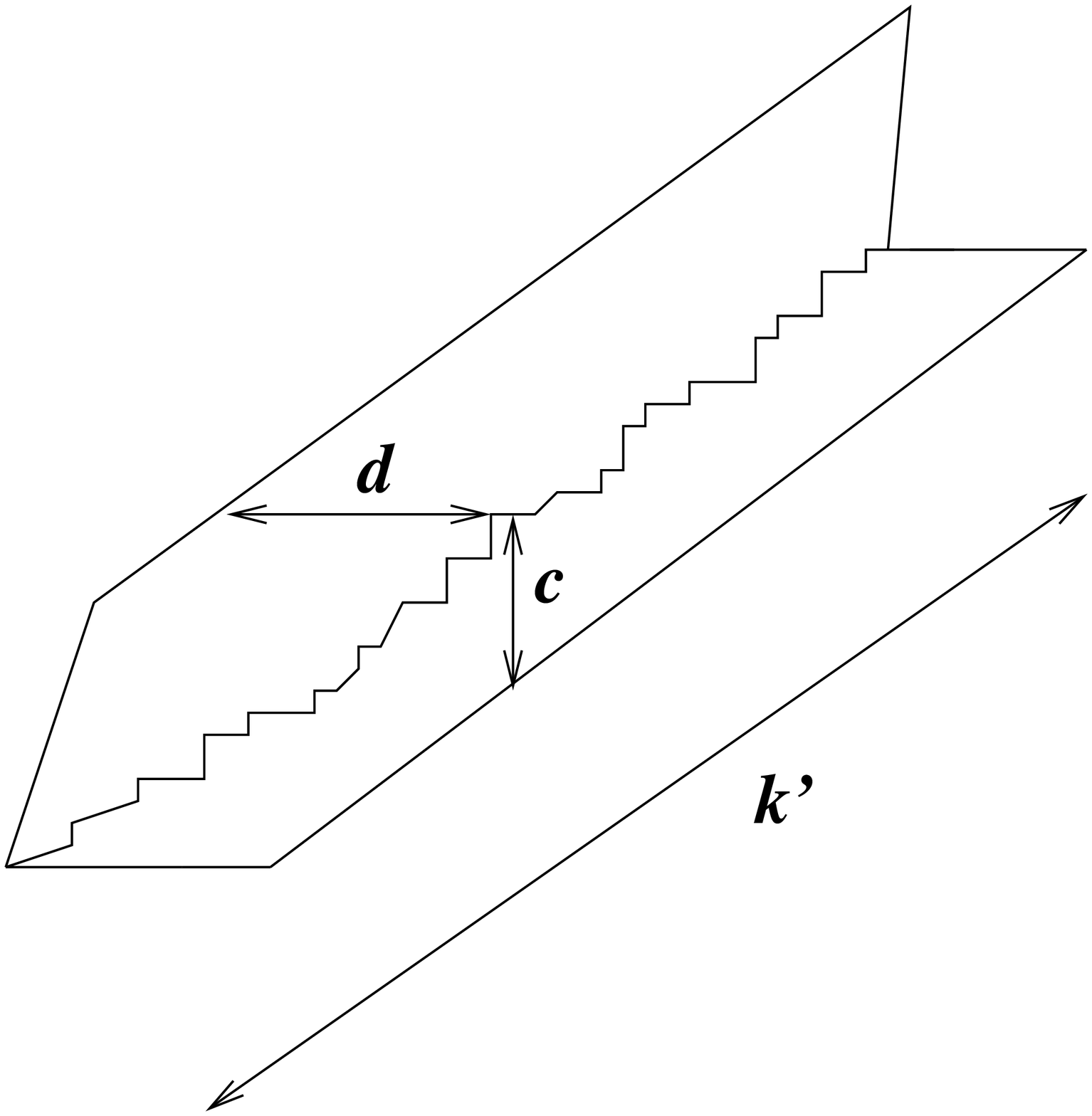}}\hfill
\vspace{-0.25in}
\caption{Typical jam configurations. For type I jams, shown in (a), $c=d=\Gamma
k'$.  For type II jams, shown in (b), $d=d_{\rm max}$ and $c=c_{\rm
max}$.  Knowing the slopes of the lines in (a), $s=1$,$s_r=2$, and
$s_b=1/2$, we can determine all of the angles:  $\phi =
\tan^{-1}(1/2)$, $\theta=\pi/4$, $\sigma = \tan^{-1}2$, and $\xi =
(\pi/2 - \tan^{-1}2)$.}
\label{fig:geom}
\end{figure}

The structure of the jams and the relevant geometric factors are shown
in Fig.~\ref{fig:geom}(a).  The jams form trapezoidal shapes of width
$\delta$. From simple geometric considerations, we can show that
$\delta = k'_i \sin(\theta-\phi) \equiv \kappa k'_i$.  The jam width
per line, $d$, and per column, $c$ is
\begin{equation}
c = d = \frac{\delta}{\cos \xi} = \frac{\sin(\theta-\phi)}{\cos \xi}
k'_i \equiv \Gamma k'_i.
\end{equation}
However, note this is only true provided that there are enough
particles available in each column and in each row.  Otherwise,
\begin{equation}
d=d_{\rm max}=\rho L/2\wr \Theta_r(\wr), {\rm \ \ \ and \ \ \ } 
c=c_{\rm max}=\rho L' \Theta_b(\wb)/2\wb.
\end{equation}

From similar geometric considerations we can show the length of the
blue jam, $l_b$, is approximately
\begin{equation}
l_b \approx b \left\{\cos \phi +
\frac{\sin \phi }{\tan(\sigma - \phi)} \right\}
\equiv b \gamma.
\end{equation}
Similarly, the length of the red jam, $l_r$, is approximately
\begin{equation}
l_r \approx a \left\{\cos \xi +
\frac{\sin \xi}{\tan\left(\sigma -
\phi\right)}\right\} \equiv a \varphi.
\end{equation}
The total number of particles involved in the jams, $J$, is the number
of jams multiplied by the width and length:
\begin{eqnarray}\label{eqn:j-rect}
J & \approx & n_j \delta \left(l_b + l_r\right) \nonumber \\
  & = & \left(2\wr L' - L\right) \kappa \left[\left(\frac{3}{2} \rho
  L/\Theta_b(\wb)\right)\gamma + \left(\frac{3}{2} \rho
  L'/\Theta_r(\wr)\right)\varphi \right] \frac{n_j}{n_k} .
\end{eqnarray}
For the data plotted in Fig.~\ref{fig:fibo-55x89}, $L'$ and
$L$ are two successive Fibonacci numbers, hence $L/L' = (1+\sqrt{5})/2
\equiv {\cal{F}}$.  In agreement with expected values shown in
Table~\ref{table:winding}, $\wr=1$ (thus $\Theta_r(\wr)=1$), and
$\wb=3$ (thus $\Theta_b(\wb)=3$).  For this geometry, $n_j=n_k=2$.
Plugging in these values into Eq.~\ref{eqn:j-rect},
\begin{equation}
J = (2 - {\cal{F}}) \kappa \left[\frac{1}{2} \gamma + \frac{3}{2{\cal{F}}}
\varphi\right] \rho L L'
\approx 0.2570 \cdot \rho L L'.
\end{equation}
Solving for the velocity, noting that the overall number of particles,
$N=\rho L L'$,
\begin{equation}\label{eqn:vr1}
v_{r1} = 1 - \frac{J}{N} \approx 0.7430.
\end{equation}
Note, this is independent of $\rho$ and independent of $L$ and $L'$.
This predicted value for $v_{r1}$ is included in the plot of the
experimentally determined velocities, shown in
Fig.~\ref{fig:fibo-55x89}.  Note the calculated value slightly
underestimates the number of particles involved in the jam (hence
slightly overestimates $v_{r1}$).

\subsubsection{Rectangular aspect ratios, type II:} As mentioned,
rectangular lattices are well described by the formalism in
Sec.~\ref{sec:geom}.  We observe the ``rich'' jam described above, but
also a second type of ``depleted'' jam (not enough
particles). Empirically, all observations of this type have $s\approx
2/3$, and one large jam.  See for instance
Fig.~(\ref{fig:fibo-55x89}c).  The per row and per column width of the
jams are the maximum, $d_{\rm max}$ and $c_{\rm max}$ respectively,
since $\Gamma k'$ is greater than the number of particles available in
a row or column.  A jam interface of length $k'$ and slope $s$ (with
$s>1$) involves $k'$ columns, but only $sk'$ rows.  Hence the total
number of particles involved in the jams:
\begin{eqnarray}
J & = & k' \left(s\cdot d_{\rm max} + c_{\rm max}\right) \nonumber \\
  & = & \left[(2\wr L' - L)/(2s-1)\right]\left[s \rho L/2\wr\Theta_r(\wr) +
  \rho L' \Theta_b(\wb)/2\wb\right].
\end{eqnarray}
For the realizations contributing to the plot in Fig.~\ref{fig:fibo-55x89},
$L/L' = (1+\sqrt{5})/2 \equiv {\cal{F}}$, $\wr=1$ (thus
$\Theta_r(\wr)=1$), $\wb=2$ (thus $\Theta_b(\wb)=2$), and $s=2/3$
(which is empirically determined).  Plugging these values in we find:
\begin{equation}
J = \frac{5}{2}\left(\frac{2}{{\cal{F}}} -
1\right)\cdot\left(\frac{7}{20}{\cal{F}} + \frac{1}{2}\right)\cdot\rho
L L' \approx 0.6293 \cdot \rho L L'. 
\end{equation}
Thus the average velocity,
\begin{equation}\label{eqn:vr2}
v_{r2} = 1 - \frac{J}{N} \approx 0.3707.
\end{equation}
We include this predicted value in the plot of
Fig.~\ref{fig:fibo-55x89}.  Note the agreement with the experimental
data. 

\subsubsection{Square aspect ratios:} The typical geometry of an
intermediate state for a square lattice is shown in
Fig.(\ref{fig:intermediate}a).  The interfaces and especially the
edges of the bands are disordered and jagged. The slope assumptions
only hold approximately: $s_r \approx 1/2$ and $s_b\approx 2$.
Furthermore, there are several particles moving freely in the low
density regions, unlike for the rectangular lattices, where all
particles eventually order into the bands and jams.  Also unlike for
the rectangular lattices (see for instance
Fig.(\ref{fig:intermediate}b)), the red and blue bands cross through
each other without the pronounced shifting upwards.

We cannot use the formalism developed above in
Sec.~\ref{subsec:winding} for this situation, since that formalism is
based solely on geometric constraints of winding seamlessly around the
lattice.  A configuration on a square lattice with $s_r=1/2$, $s_b=2$,
$s=1$, would have $w_r=1/2$. Plugging into Eqn.~(\ref{eqn:kprime}) we
find the required length of the overall interface, $k'=0$.  Instead,
empirically we find the blue jams form a trapezoidal shape of
approximate length $b$ and height $a/4$.  Likewise the red jams form a
trapezoid of approximate length $a$ and width $b/4$.  Each jam has
this shape and there are $n_j=3$ jams altogether ({\em i.e.}, three
distinct intersections of the bands). The number of particles involved
in jams, $J$ is:
\begin{equation}
J \approx n_j (a b/4 + a b/4) = 3 a^2/2.
\end{equation}
Using Eq.~\ref{eqn:a} to solve for $a$ and the fact that the overall
number of particles, $N=\rho L L'$, we can solve for the fraction of
particles in the jammed state:
\begin{equation}
\frac{J}{N} \approx \frac{3}{2} \left( \frac{3}{4} \rho L
\right)^2 \frac{1}{\rho L^2} = \frac{27}{32}\rho.
\end{equation}
Hence the velocity, 
\begin{equation}\label{eqn:v_s}
v_s \approx 1 - \frac{J}{N} = 1 - \frac{27}{32}\rho.
\end{equation}
This predicted value $v_s$ is included in the plots of
Fig.~\ref{fig:bml-64-512}.  It captures the features of the
experimental data, including the slight decrease in $v_s$ with
increasing $\rho$.

\section{Discussion}\label{sec:disc}

The BML traffic model is a simple model of a jamming transition with
self organization.  In our study, instead of agreement with
conventional beliefs, we find stable intermediate configurations with
phase coexistence of jammed and free-flowing traffic.  Such
configurations have not been previously reported in the literature,
despite the extensive amount of past work on the BML model.
Furthermore, these intermediate configurations have interesting
geometric and topological properties, with different behaviors
resulting as a consequence of different aspect ratios of the
underlying lattice.  We develop a formalism, based on geometric
constraints imposed by the lattice, to predict the asymptotic
velocities of the coexisting phases.  Visualizing the kinetic pathways
of the evolving configurations was a key element in uncovering the
existence of the intermediate phases and, moreover, their periodic
nature on lattices with relatively prime aspect ratios.  The
observations described in this manuscript open up a range of new
questions about the BML model.

As mentioned, instead of a phase transition as a function of density,
we observe a bifurcation point where the intermediate states first
begin appearing, and a second bifurcation point, where they completely
cease to appear.  Perhaps more interesting than predicting the
asymptotic velocities, would be to calculate the locations of the
bifurcation points.  From our experimental data, the exact location of
the bifurcation points are difficult to determine, and moreover, also
depend on the aspect ratio of the underlying lattice.

It is possible that there is a sharp phase transition. However, in
such a case, the density $\rho$ would not be the appropriate order
parameter.  Perhaps a more appropriate order parameter would be an
interaction energy between north-bound and east-bound particles.  Note
that when in the free-flowing state, the north and east particles have
moved onto non-interacting lattices.  It may be possible that one can
define an initial energy based on the overlap or interaction between
two lattices, and use that as an order parameter.

A complication which makes theoretical treatment of the BML model
difficult, is that it is not strictly monotonic. Adding particles to a
configuration that is known to jam ({\em i.e.}, increasing $\rho$),
can actually change the sequence of particle interactions and result
in that configuration going to free-flowing instead of jamming.
Furthermore, it is known that certain discrete models with the same
property as BML---namely that the randomness is in the initial
condition, yet the dynamics fully deterministic---can be notoriously
difficult to deal with analytically. Examples include bootstrap
percolation\cite{Holroyd-bootstrap} and the Lorentz lattice
gas\cite{grimmett-pinball}.  We modified the BML model to include a
small probability for particles to flip species-types at each update.
Our preliminary studies, adding this small amount of randomness to the
dynamics, suggest that the model with randomness has extremely
different geometric properties than the original BML model.  In
addition, for the model with randomness, we did not observe the
intermediate configurations described herein.

\acknowledgments{This work has benefited greatly from discussions with
L\'{a}szl\'{o} Lov\'{a}sz, Alexander Holroyd, and Roman Kotecky.}

\bibliographystyle{unsrt}
\bibliography{/home/raissa/Bibl/percolation,%
/home/raissa/Bibl/cars,%
/home/raissa/Bibl/ca,%
/home/raissa/Bibl/markov,%
/home/raissa/Bibl/statistics,%
/home/raissa/Bibl/rwalks}

\end{document}